\documentclass{jpp}

\usepackage{graphicx}
\usepackage{natbib}

\ifCUPmtlplainloaded \else
  \checkfont{eurm10}
  \iffontfound
    \IfFileExists{upmath.sty}
      {\typeout{^^JFound AMS Euler Roman fonts on the system,
                   using the 'upmath' package.^^J}%
       \usepackage{upmath}}
      {\typeout{^^JFound AMS Euler Roman fonts on the system, but you
                   dont seem to have the}%
       \typeout{'upmath' package installed. JPP.cls can take advantage
                 of these fonts, if you use 'upmath' package.^^J}%
      }
  \else
  \fi
\fi

\ifCUPmtlplainloaded \else
  \checkfont{msam10}
  \iffontfound
    \IfFileExists{amssymb.sty}
      {\typeout{^^JFound AMS Symbol fonts on the system, using the
                'amssymb' package.^^J}%
       \usepackage{amssymb}%

      }{}
  \fi
\fi

\ifCUPmtlplainloaded \else
  \IfFileExists{amsbsy.sty}
    {\typeout{^^JFound the 'amsbsy' package on the system, using it.^^J}%
     \usepackage{amsbsy}}
    {\providecommand\boldsymbol[1]{\mbox{\boldmath $##1$}}}
\fi

\newsavebox{\astrutbox}
\sbox{\astrutbox}{\rule[-5pt]{0pt}{20pt}}

\title[On the Value of the Reconnection Rate]{On the Value of the Reconnection Rate}

\author[L. Comisso and A. Bhattacharjee]%
{L.\ns C\ls O\ls M\ls I\ls S\ls S\ls O  \ns
\and A.\ns B\ls H\ls A\ls T\ls T\ls A\ls C\ls H\ls A\ls R\ls J\ls E\ls E}

\affiliation{Department of Astrophysical Sciences and Princeton Plasma Physics Laboratory,\\ Princeton University, Princeton, NJ 08544, USA}

\begin{document}

\maketitle

\begin{abstract}
Numerical simulations have consistently shown that the reconnection rate in certain collisionless regimes can be fast, on the order of $0.1 {v_A}{B_u}$, where $v_A$ and ${B_u}$ are the  Alfv{\'e}n speed and the reconnecting magnetic field upstream of the ion diffusion region. This particular value has been reported in myriad numerical simulations under disparate conditions. However, despite decades of research, the reasons underpinning this specific value remain mysterious. Here, we present an overview of this problem and discuss the conditions under which the ``0.1 value'' is attained. Furthermore, we explain why this problem should be interpreted in terms of the ion diffusion region length.
\end{abstract}

\section{Introduction with Overview of the Problem}

Magnetic reconnection is a fundamental plasma process that occurs in a wide variety of laboratory, space, and astrophysical plasmas. Its definition is meaningful in plasmas that are almost ideal, i.e., in those cases where magnetic field lines ``move'' with the plasma in the vast majority of the domain, while the breaking of the magnetic field line connectivity occurs only in very localized diffusion regions. This reconnection process can enable a rapid conversion of magnetic energy into thermal, supra-thermal, and bulk kinetic energy. As such, magnetic reconnection is believed to play a key role in many of the most striking and energetic phenomena such as sawtooth crashes, magnetospheric substorms, coronal mass ejections, stellar and gamma-ray flares \citep[]{Tajima1997,Kulsrud2005,Yamada2010}.

In order to explain the magnetic energy conversion rates associated with these phenomena, it is essential to know the rate at which magnetic reconnection occurs. The reconnection rate quantifies the temporal rate of change of magnetic flux that undergoes the reconnection process. 
When the system under consideration is translationally invariant in one direction, the reconnection rate can be expressed as 
\begin{equation}\label{}
\frac{{d\Phi}}{{dt}} = \frac{d}{{dt}}\int_S {{\vec B} \cdot d{\vec S}}  = \oint_{\partial S} {\vec E \cdot d\vec l} = \int_{X{\rm{-line}}} {{E_z}dl}   \, .    
\end{equation}
Here, $\Phi$ is the magnetic flux through the surface $S$ bounded by the contour $\partial S$ encompassing the $X$-line. An $X$-line is the projection of an hyperbolic point for the magnetic field along the ignorable direction. Therefore, the reconnection rate is a measure of the rate at which magnetic flux is transported across the $X$-line. In a more general three-dimensional case, the evaluation of the reconnection rate is more subtle. A general approach \citep[]{Hesse2005} would be to quantify the reconnection rate as 
\begin{equation}\label{Rec_Hesse}
\frac{{d\Phi}}{{dt}} = \max \left( {\int {{E_\parallel }ds} } \right) \, ,
\end{equation}
where $s$ represents the parametrization of the magnetic field lines, and the integral has to be performed over all field lines passing through the non-ideal region (where $E_\parallel = \vec E \cdot \vec B/ | {\vec B} | \neq 0$). The measure (\ref{Rec_Hesse}) is an attractive choice for quantifying the reconnection rate, but there are some caveats associated with it. Indeed, there could be some ambiguity related to the field line integration of $E_\parallel$, as in regions where magnetic field lines are stochastic \citep[]{Borgogno2005}, or it may be not possible to distinguish between reconnection and simple diffusion \citep[]{Huang2014}. In addition, this measure can be applied only in the presence of a non-vanishing magnetic field. If this is not the case, the reconnection rate may be calculated by combining the line integrals of $E_\parallel$ along magnetic separators 
\citep[]{Greene1988,LauFinn1990,Wilmot2011}, 
which are magnetic field lines connecting two null points (i.e., points at which $|{\vec B}|=0$). As this brief discussion may suggest, a completely general and practical measure of the reconnection rate is still lacking, and indeed it constitutes an important ongoing area of research (see, for example, the discussion given by \citet[]{Dorelli2008} in the context of the Earth's magnetosphere.)

The problem of determining the reconnection rate of a magnetic reconnection process dates back to the 1950's. At that time, the astrophysical community was trying to understand if magnetic reconnection could have served as the mechanism underlying solar flares, which are bursts of high-energy radiation from the Sun's atmosphere that strongly affect the space weather surrounding the Earth. A simple resistive magnetohydrodynamic (MHD) model of magnetic field line merging was proposed by \citet[]{Sweet}, and then, with the contribution of \citet[]{Parker}, the reconnection rate was evaluated. They considered a quasi-stationary reconnection process occurring within a two-dimensional current sheet. Then, assuming an incompressible flow, the normalized reconnection rate (per unit length) can be shown to be
\begin{equation}\label{eq1}
\frac{1}{{{v_A}B_{u}}} \frac{{d{\Phi}}}{{dt}} \sim S^{-1/2} (1 + P_m)^{1/4}  \, .   
\end{equation}
In this formula, $S := v_A L/\eta$ and $P_m := \nu/\eta$ are the Lundquist number and the magnetic Prandtl number, respectively. As usual, $\eta$ indicates the magnetic diffusivity and $\nu$ the kinematic viscosity. The Lundquist number is evaluated using the current sheet half-length $L$ and the Alfv{\'e}n speed $v_A = B_u {\left( {{\mu _0}\rho } \right)^{-1/2}}$, where $B_u$ is the reversing magnetic field upstream of the current sheet. In reality, Eq. (\ref{eq1}) is not exactly the Sweet-Parker formula for the reconnection rate, but represents its generalization to account for plasma viscosity \citep[]{Park1984}. 

The Sweet-Parker model of reconnection is faster than simple diffusion, but for very large $S$ systems, such as those found in most space and astrophysical environments, it is far too slow to explain the observed fast energy release rates. In order to bypass this limitation, \citet[]{Petschek1964} proposed a different model in which a relatively short reconnection layer acts as a source for two pairs of slow mode shocks, allowing for much faster reconnection rates. This model was subsequently generalized by \citet[]{Priest1986}, who put forward a wider family of ``almost-uniform models'' that include Petschek's model as a special case. However, these models have not been supported by numerical simulations \citep[]{Biskamp1986}, which have shown that Petschek-like configurations cannot be sustained in the context of MHD with constant resistivity. Petschek's mechanism can occur within the resistive-MHD framework if the plasma resistivity increases sharply in the reconnection layer \citep[]{Kulsrud2001,Kulsrud2011}, but the difficulties in firmly establishing the nature and details of such anomalous resistivity have led the scientific community to look for other alternatives. 

An important advance occurred when \citet[]{BHYR_2009}, and later \citet[]{Cassak2009}, showed that the predictions of the Sweet-Parker model break down for large $S$ values because of the occurrence of the plasmoid instability \citep[]{Biskamp1986,Tajima1997,Loureiro2007,Comisso2016,Comisso2016B}. In the high-Lundquist number regime, the reconnection process in the nonlinear regime becomes strongly time-dependent due to the continuous formation, merging, and ejection of plasmoids. An estimation of the time-averaged reconnection rate in this regime was proposed by \citet[]{Huang2010}, as well as by \citet[]{Uzdensky2010}, and it has been generalized to account for plasma viscosity as \citep[]{Comisso2015,Comisso2016} 
\begin{equation}\label{eq2}
\frac{1}{{{v_A}{B_u}}}\left\langle {\frac{{d{\Phi}}}{{dt}}} \right\rangle \sim 10^{-2} {(1 + P_m)^{-1/2}} \, ,   
\end{equation}
where $\left\langle  \ldots  \right\rangle$ denotes time-average. This formula shows that, for high-Lundquist numbers, the (time-averaged) reconnection rate becomes independent of the Lundquist number (but not the magnetic Prandtl number) and much higher than the Sweet-Parker rate for very large $S$-values.

Other MHD models of reconnection have also been investigated. In particular, since the pioneering work by \citet[]{Matthaeus1986}, turbulence effects have been shown to produce a distribution of reconnection sites that is capable of increasing the global reconnection rate \citep[]{Servidio2009}. The impact of turbulence and the plasmoid instability on the reconnection rate has caused a rethinking of magnetic reconnection in MHD plasmas. However, in many situations the current layers that form reach scales at which two fluid/kinetic effects become important. In all these cases, an MHD description fails to reproduce accurately the physics of the reconnection process, and two fluid and kinetic effects must be considered.

For the aforementioned reasons, a complementary path in investigating fast magnetic reconnection has been pursued at least since the 1990's by means of numerical simulations of Hall MHD, two-fluid and kinetic models. Several research groups have shown that collisionless effects were able to strongly speed up the reconnection process \citep[]{Aydemir1992,OttPor1993,Wang1993,Mandt1994,Biskamp1995,Kleva1995,MaBhatta1996,Shay1999,Grasso1999,Birn_2001,Porcelli2002}. In particular, numerical simulations consistently demonstrated that the reconnection rate in certain collisionless regimes becomes
\begin{equation}\label{eq3}
\frac{1}{{{v_A}B_{u}}} \frac{{d{\Phi}}}{{dt}} \sim 0.1  \, ,
\end{equation}
a value that is compatible with many observations and experiments \citep[]{Yamada2010}, meaning that collisionless effects may be crucial to explain many magnetic reconnection phenomena. 
Note that even here (and in the following) we have considered the reconnection rate per unit length in the out-of-plane direction, whereas $v_A$ and $B_u$ are evaluated upstream of the ion diffusion region, which can be seen as the region where ${E_z} + {({{\vec v}_i} \times \vec B)_z} \ne 0$. Although the relation (\ref{eq3}) was found to be valid only in the steady-state limit, or in the vicinity of the peak reconnection rate, it was nevertheless surprising to discover that ${({v_A}{B_u})^{ - 1}}d\Phi /dt$ seemed to be unaffected by the microphysics and macrophysics of specific models. This intriguing result led \citet[]{Shay1999} to speculate that the aforementioned value could be universal. Such a conjecture stimulated a long debate in the plasma physics and astrophysics communities - one that continues to this day. \emph{Is the reconnection rate value of $0.1$ truly universal?} \emph{What are the physical reasons of this particular value?}

In order to explain the fast reconnection rates observed in numerical simulations, \citet[]{Shay1999} brought forward an argument by \citet[]{Mandt1994}, who proposed that fast magnetic reconnection is enabled by the presence of fast dispersive waves. These waves would speed up the reconnection process by giving rise to the development of a Petschek-type outflow configuration. In contrast, the absence of dispersive waves would lead to an extended Sweet-Parker-type layer, forcing collisionless reconnection to be slow in large systems \citep[]{Rogers2001}. This argument, however, was found not to be true. Indeed, numerical simulations have shown that fast magnetic reconnection also occurs in electron-positron plasmas, which do not support fast dispersive waves \citep[]{Bessho2005,DaugKar2007,Chacon2008,Zenitani2008}. More recently, \citet[]{Liu_2014}, as well as \citet[]{Stanier2015}, have reconsidered this argument and have shown that, in an electron-ion plasma, fast reconnection is also manifested in the strongly magnetized limit (where fast dispersive waves are suppressed) defined by $\beta := 2{\mu_0}{n_0}{k_B}(T_e + T_i)/B^2 \ll {m_e}/{m_i}$ and $B_{u}^2 \ll ({m_e}/{m_i}){B^2}$.

While several works have shown that fast dispersive waves are not required for fast magnetic reconnection, they 
have also confirmed that the maximum/steady-state reconnection rate satisfies Eq. (\ref{eq1}) \citep[e.g.,][]{DaugKar2007,Liu_2014,Stanier2015}. There are also some works that have argued against the $\sim 0.1$ value of the maximum/steady-state reconnection rate \citep[e.g.,][]{Porcelli2002,Fitzpatrick2004,Bhatta2005,Andres2016}. 
In light of the subtlety of the problem, we shall elucidate the conditions under which one should expect a maximum/steady-state reconnection rate $\sim 0.1$. We will also present some thoughts on this apparent commonality of the reconnection rate, which still remain a mystery, and constitutes an important unsolved problem in magnetic reconnection theory. We refer to this problem as the ``$0.1$ problem''.

\section{Thoughts on the interpretation of this problem} \label{sec:solve_problem}

Hitherto, we have focused on summarizing some important discoveries and ideas that underlie the reconnection rate and the $0.1$ problem. In this section, we will present some thoughts as to how to interpret this problem.

The first important point that cannot be overlooked is that {\it not} all of the collisionless reconnection processes give rise to a peak/steady-state reconnection rate $\sim 0.1$. Indeed, this value is attained only if the system under consideration is {\it strongly unstable} (e.g., the tearing stability parameter $\Delta '$ is greater than a certain threshold) and/or {\it forced} (e.g., the externally imposed flow or magnetic perturbation exceeds a certain threshold). In the following, we will assume that this is the case. Otherwise, the reconnection rate can be arbitrarily low (e.g., Rutherford-like evolution).

In principle, there are no fundamental reasons to believe that different physical models - e.g., Hall-MHD, extended-MHD,  multi-fluid, gyrofluid, hybrid, gyrokinetic, kinetic - which are characterized by different physics at the $X$-line, should yield the same peak/steady-state reconnection rate. Indeed, one may think that the commonality of the $\sim 0.1$ value is just a coincidence.
However, 
if the same motif is repeated many times, that cannot be coincidence \citep[]{Christie1936}, especially when one considers all the differences inherent in the many numerical simulations that have reported this reconnection rate. In particular, it appears that:
\begin{enumerate}
\item The steady-state reconnection rate is not linked to the microphysics of the electron diffusion region \citep[e.g.,][]{Shay1998,Stanier2015}.
\item The $\sim 0.1$ value of the reconnection rate is independent of the system size \citep[e.g.,][]{Shay2004,Comisso2013}.
\item The $\sim 0.1$ value occurs even when the field structures (e.g., current density) are very different \citep[e.g.,][]{Liu_2014,Stanier2015}.
\item 3D simulations, despite exhibiting great differences in the structure of the reconnection layer, give reconnection rates similar to those of 2D simulations \citep[e.g.,][]{Daughton2014,Guo2015}.
\item Simulations in turbulent scenarios lead to current sheets characterized by the same reconnection rate as in the standard laminar picture \citep[e.g.,][]{Wendel2013,Daughton2014}.
\end{enumerate}

To correctly interpret the aforementioned results, we argue that is necessary to shift the focus from the reconnection rate itself, and we conjecture that the reconnection rate is actually {\it not} the real ``universal quantity'', but it is derived from a more fundamental one, the {\it aspect ratio of the ion diffusion region} $\Delta /L$. It is not the former that is $\sim 0.1$, but it is the latter which takes on this value. Then, from mass conservation in steady state, $\vec \nabla  \cdot (n {\vec v}) = 0$, one obtains
\begin{equation}\label{eq4}
\frac{{{v_{in}}}}{{{v_{out}}}} \frac{{{n_{in}}}}{{{n_{out}}}} \sim \frac{\Delta}{L} \sim 0.1  \, .
\end{equation}
It is only when the flow is incompressible, $\vec \nabla  \cdot {\vec v} = 0$, and the outflow velocity is $v_{out} \sim v_A$, that the reconnection rate turns out to be $\sim 0.1$. 
The discrepancy between the aspect ratio and the reconnection rate can become particularly evident when considering magnetic reconnection in the relativistic regime. Indeed, the inflow velocity may increase due to Lorentz contraction as per the relation ${v_{in}}/{v_{out}}  \sim  ({\gamma _{out}}{n_{out}}/{\gamma_{in}}{n_{in}})\Delta /L$, where $n$ is the proper particle number density and $\gamma := {(1 - {v^2}/{c^2})^{ - 1/2}}$. 
This reconnection rate enhancement has been clearly found in electron-positron plasmas, which are characterized by $\Delta \sim L [ S^{-1}(1 + {P_m})^{1/2} + H^{-1}]^{1/2}$, where 
$H := f^{-1} (2L/\lambda_e)^2$ 
is the thermal-inertial number \citep[]{Comisso2014} and $f=K_3(\zeta)/K_2(\zeta)$ is the relativistic thermal factor ($K_n$ indicates the modified Bessel function of the second kind of order $n$ and $\zeta$ defines the ratio of rest-mass energy to thermal energy).  Indeed, in the strictly collisionless regime, \citet[]{Liu2015} have performed kinetic simulations that have shown an enhancement of ${v_{in}}/{v_{out}}$ consistent with the increase of ${\gamma_{in}}/{\gamma_{out}}$.


There is another important point that needs to be considered. This involves an essential difference between collisional and collisionless reconnection. In collisional MHD models, the thickness of the diffusion region depends on its length $L$, precisely 
\begin{equation}\label{}
\frac{\Delta}{L^{1/2}}  \sim {\left( {\frac{\eta}{v_A}} \right)^{1/2}} {(1 + {P_m})^{1/4}} \, .     
\end{equation}
On the other hand, in two-fluid/kinetic collisionless models, the width of the ion diffusion region depends only on the details of the microphysics of the reconnection process:
\begin{equation}\label{}
\Delta \sim {\rho _{se}} = \frac{{{c_{se}}}}{{{\omega_{ci}}}}, \, \, {\rho_{s}} = \frac{{{c_{s}}}}{{{\omega _{ci}}}}, \, \, {\lambda_i} = \frac{c}{\omega_{pi}}, \, \, {\lambda_e} = \frac{c}{\omega_{pe}}, \, \, ....     
\end{equation}
This width is associated to very different length scales, such as the cold ion sound Larmor radius ${\rho _{se}}$, the ion sound Larmor radius ${\rho_{s}}$, the ion skin depth $\lambda_i$, the positron/electron skin depth $\lambda_e$, etc. Therefore, if $\Delta/L \sim 0.1$ holds true for all the different two-fluid/kinetic models, it means that $L$ self-adjust itself in such a way to match the observed aspect ratio.

Given the above refocusing observations, the key question of the $0.1$ problem shifts to: \emph{why does the ion diffusion region self-adjust in such a way that the length obeys $L \sim 10 \Delta$?}

Here, without intending to furnish a solution to this problem, we provide qualitative arguments to illustrate why $L$ cannot be significantly different from $10 \Delta$. 
This can be shown heuristically in the following manner. Let us start by assuming $L = \Delta$ and examine its validity. This limit has been studied extensively in the past \citep[e.g.,][]{Priest1985,Priest2000} and is relevant to the solution of the $0.1$ problem (e.g., P.A. Cassak and M.A. Shay, private communication, 2016). 
In this instance, it is possible to demonstrate that the reconnection rate vanishes in a plasma. To this purpose, one can exploit the symmetries of the problem, which dictate that
\begin{equation}\label{}
{v_x}( \pm x, \mp y) =  \pm {v_x}(x,y) \, , \quad {v_y}( \pm x, \mp y) =  \mp {v_y}(x,y) \, ,
\end{equation}
\begin{equation}\label{}
{B_x}( \pm x, \mp y) =  \mp {B_x}(x,y) \, , \quad {B_y}( \pm x, \mp y) =  \pm {B_y}(x,y) \, ,
\end{equation}
if the reconnection occurs at a symmetric $X$-point configuration, with the $X$-point situated at the origin. Combining the above properties with the fact that the reconnection rate for $L = \Delta$ must be invariant under a point reflection of the velocity field
\begin{equation}\label{}
({v_x}, {v_y}, {B_x}, {B_y}) \mapsto (-{v_x}, -{v_y}, {B_x}, {B_y}) \, ,
\end{equation}
it follows that the only possible solution is $v_{in} = v_{out} = 0$. This implies that the reconnection process chokes itself off if $L = \Delta$.
It is straightforward to check that a steady reconnection process is also not possible for $L < \Delta$. Indeed, in this case, the current density at the $X$-point would act to decrease the inflow velocity.

Next, we consider the case $L > \xi \Delta$, where $\xi \gg 1$ represents a coefficient that will be discussed soon. If a current sheet remains stable over time for arbitrarily large $\xi$, the possibility of obtaining a fast reconnection rate $\sim 0.1$ would be precluded. 
However, extended current sheets are subject to a tearing-like (plasmoid) instability \citep[e.g.,][]{Biskamp1986}.
This implies that for $\xi > \xi_c$, with $\xi_c$ indicating a critical threshold value, the global current sheet breaks up and is replaced by a chain of plasmoids/flux ropes of different sizes separated by smaller current sheets \citep[]{Shibata2001}.

In a reconnection layer dominated by the presence of plasmoids, the complexity of the dynamics gives rise to a strongly time-dependent process \citep[e.g.,][]{Daughton2009}. Nevertheless, if this process can reach a statistical steady state, we may expect that the current sheet located at the main $X$-point, which is the one that determines the global reconnection rate, should not be longer than the marginally stable sheet \citep[]{Huang2010, Uzdensky2010, Comisso2015, Comisso2016}.
Indeed, the fractal cascade arising from the plasmoid instability terminates when the length of the innermost local current layer is shorter than the critical length $L_c = \xi_c \Delta$. The local current sheet situated at the primary $X$-point could be subjected to continual stretching by plasmoids moving in the outflow direction, but the plasmoid instability occurs if its length exceeds $L_c$. Thus, it is reasonable to assume that the length of the local current sheet at the main $X$-point should not exceed $L_c$.
As a consequence, the length of the main diffusion region remains bounded from above, but a clear-cut value of $\xi_c$ remains unknown. Although at present there are no analytical estimates of the aspect ratio of the main diffusion region, numerical simulations have found $\xi_c \sim 50$ in the collisionless regime \citep[]{Daughton2006,Ji2011}.


According to the above arguments, it is clear that the length of the ion diffusion region that determines the reconnection rate is bounded from above and below as $\Delta  < L \lesssim 50 \Delta$ in a quasi-steady (or statistical steady state) strongly driven/unstable collisionless reconnection process. 

\section{Final remarks}

We have seen that the maximum/steady-state reconnection rate is regulated by the length of the ion diffusion region. However, so far we have not stressed the importance of the boundary conditions on the diffusion region length.
Boundary condition may indeed have a strong impact on the length of the current sheets if the computational domain is not sufficiently large. Therefore, the choice of the boundary conditions require extra caution. For example, periodic boundary conditions may force the length of a current sheet to remain small, limiting the duration in which the results are physically meaningful \citep[]{Daughton2006}.

The knowledge of the maximum/steady-state reconnection rate is crucial when trying to understand whether magnetic reconnection can be fast enough to account for the energy release time-scales observed in a specific system. 
This is because most of the magnetic flux reconnection takes place during this stage of the process.
It is therefore not surprising that much of the magnetic reconnection research done to date has focused on this issue. However, we wish to end our discussion by noting that there are other important questions that lie beyond the paradigm of the maximum/steady-state reconnection rate. While this observable could be insensitive to many features of the specific model, the reconnection rate evolution is not. Indeed, it can be extremely different in diverse systems, since the initial evolution of any reconnection process depends on the details of the microphysics as well as the large-scale ideal-MHD conditions. This initial (typically linear) stage could be completely negligible in terms of magnetic flux reconnection, but it is crucial for determining whether a particular system has enough time to accumulate the magnetic energy that is mostly liberated during the faster stage of the reconnection process. This issue, which is commonly referred to as the onset problem, is also an important and active area of research.



\section*{Acknowledgments}

It is a pleasure to acknowledge the fruitful and lively discussions held during the symposium ``Solved and Unsolved Problems in Plasma Physics'', held in celebration of the career of Professor Nathaniel Fisch, who has made an art form of asking simple but deep questions that have led to remarkable discoveries. We are indebted to Russell Kulsrud for many enlightening discussions and his suggestion to exploit the symmetries of the problem to construct an elegant proof that steady-state reconnection cannot occur for $L = \Delta$. We are particularly grateful to Manasvi Lingam and Felipe Asenjo, who read the manuscript and provided important suggestions. Finally, we would like to acknowledge stimulating discussions with Paul Cassak, William Daughton, Will Fox, Daniela Grasso, Yi-Min Huang, Hantao Ji, Yi-Hsin Liu, Jonathan Ng, Adam Stanier and Masaaki Yamada. 
This research is supported by the NSF Grant Nos. AGS-1338944, AGS-1552142, and the DOE Grant No. DE-AC02-09CH-11466.




\end{document}